\documentclass[11pt]{article}

\begin{document}

\title{An Instability in a Straightening Chain}

\author{J. A. Hanna\thanks{hanna@physics.umass.edu}, H. King \\
Department of Physics, University of Massachusetts, Amherst, MA 01003}

\date{\today}

\maketitle

\begin{abstract}
We briefly discuss an arch-like structure that forms and grows during the rapid straightening of a chain lying on a table.  This short note accompanies a fluid dynamics video submission (V029) to the APS DFD Gallery of Fluid Motion 2011.
\end{abstract}

The accompanying video shows fifty feet of chain (silver-plated base metal, $\sim$2 mm links, {\footnotesize{\texttt{www.jewelrysupply.com}}}) arranged on about two feet of table top; off camera, a power drill pulls one end at a velocity $T \sim$ 8 m/s.

The relevant forces are those due to the chain's inertia, and a line tension whose role is identical to that of pressure in a one-dimensional incompressible fluid.  Gradients in this tension amplify and advect small of out-of-plane disturbances, presumably with some upward rectification by the table.  The advection velocity drops to zero as the chain approaches its terminal velocity $T$.  There, the tension is uniform, structures are nearly ``frozen'' in the laboratory frame, and the vertical slack accumulates behind and within a slowly growing arch.

By the end of the video, this arch has grown to about ten centimeters high.  However, an analysis that takes gravity into account suggests that such a structure may remain stable when grown to a height on the order of a meter, the natural length scale of the system with gravity included ($T^2 / g$).  Structures of such a size are difficult to access experimentally, as they do not evolve within the corresponding natural time scale of about one second ($T / g$).  This is because the system is oblivious to gravity during the observation period of the experiment; in a region with local curvature $\kappa$ on the order of a few tens of inverse meters, an element of chain sees a large ratio of inertial to gravitational accelerations $T^2 \kappa / g$.  

\section*{Acknowledgments}

We thank N. Menon and C. D. Santangelo for support and discussions, and acknowledge funding from National Science Foundation grants DMR 0907245 and DMR 0846582.

\end{document}